\documentclass[conference]{IEEEtran}
\IEEEoverridecommandlockouts
\usepackage[english]{babel}
\usepackage{amsmath,amssymb,amsfonts}
\usepackage{algorithmic}
\usepackage{flushend}
\usepackage{graphicx}
\usepackage{textcomp}
\usepackage{xcolor}
\usepackage{tikz}
\usepackage[caption=false]{subfig}
\usepackage{stfloats}
\usepackage[normalem]{ulem}
\usepackage{url}            % simple URL typesetting
\def\BibTeX{{\rm B\kern-.05em{\sc i\kern-.025em b}\kern-.08em
    T\kern-.1667em\lower.7ex\hbox{E}\kern-.125emX}}
    
\usetikzlibrary{intersections}
\usetikzlibrary{shapes.arrows}
\usetikzlibrary{automata,positioning}
\tikzset{MyArrow/.style={single arrow, draw=black, minimum width=10mm, minimum height=30mm, inner sep=0mm, single arrow head extend=1mm}}
\tikzset{MyArrow2/.style={double arrow, draw=black, minimum width=10mm, minimum height=30mm, inner sep=0mm, double arrow head extend=1mm}}

\title{A Lower Bound on Latency Spikes for Capacity-Seeking Network Traffic}
\makeatletter
\newcommand{\linebreakand}{%
  \end{@IEEEauthorhalign}
  \hfill\mbox{}\par
  \mbox{}\hfill\begin{@IEEEauthorhalign}
}
\makeatother
\author{\IEEEauthorblockN{1\textsuperscript{st} Bj\o rn Ivar Teigen}
\IEEEauthorblockA{
\textit{Domos}\\
Oslo, Norway \\
bjorn@domos.no}
\and
\IEEEauthorblockN{2\textsuperscript{nd} Neil Davies}
\IEEEauthorblockA{
\textit{Predictable Network Solutions Limited}\\
Stonehouse, United Kingdom \\
neil.davies@pnsol.com}
\and
\IEEEauthorblockN{3\textsuperscript{rd} Kai Olav Ellefsen}
\IEEEauthorblockA{\textit{Department of Informatics} \\
\textit{University of Oslo}\\
Oslo, Norway \\
kaiolae@ifi.uio.no}
\linebreakand
\IEEEauthorblockN{4\textsuperscript{th} Tor Skeie}
\IEEEauthorblockA{\textit{Department of Informatics} \\
\textit{University of Oslo}\\
Oslo, Norway \\
tskeie@ifi.uio.no}
\and
\IEEEauthorblockN{5\textsuperscript{th} Carlo Augusto Grazia}
\IEEEauthorblockA{\textit{Department of Engineering Enzo Ferrari} \\
\textit{University of Modena and Reggio Emilia}\\
Modena, Italy \\
carloaugusto.grazia@unimore.it}
\and
\IEEEauthorblockN{6\textsuperscript{th} Jim Torresen}
\IEEEauthorblockA{\textit{Department of Informatics} \\
\textit{University of Oslo}\\
Oslo, Norway \\
jimtoer@ifi.uio.no}
}

\begin{document}
\maketitle

\begin{abstract}
Most Internet traffic is carried by capacity-seeking protocols such as TCP and QUIC. Capacity-seeking protocols probe to find the maximum available throughput from sender to receiver, and, once they converge, attempt to keep sending traffic at this maximum rate. Achieving reliable low latency with capacity-seeking end-to-end methods is not yet entirely solved. We contribute a theoretical analysis to this ongoing discussion. In this work, we derive an expression for the minimum size of the spike in latency caused by a sudden drop in network capacity. Our results highlight a quantifiable and fundamental constraint on capacity-seeking network traffic. When end-to-end capacity is suddenly reduced, capacity-seeking traffic inevitably produces a latency spike. A lower bound on this latency spike can be calculated by multiplying the round-trip delay from the network bottleneck to the source of capacity-seeking traffic by the magnitude of the end-to-end capacity reduction. Testbed experiments show that this bound holds for the DCTCP, BBR, and Cubic congestion control algorithms. Our results have implications for the design of low-latency PHY and MAC-layer technologies because we quantify an important transport-layer consequence of unstable traffic rates.

\end{abstract}
% keywords can be removed

\section{Introduction}
\label{sec:introduction}

End-to-end congestion control methods, such as those used by TCP and QUIC, are the main ways of avoiding congestion collapse in the Internet. Most congestion control (CC) algorithms are \textit{capacity-seeking}, meaning they aim to utilize close to 100\% of the available end-to-end capacity. When capacity-seeking traffic shares a queue with latency-sensitive traffic, or when the traffic is itself latency-sensitive, the queuing delay induced by capacity-seeking traffic can potentially impair the user experience. Some applications, such as cloud gaming, can require both high throughput and very low end-to-end latency to maintain a good user experience \cite{Chen2014OnSystems}. Cloud gaming applications typically use a capacity-seeking protocol to adapt the video resolution \cite{Graff2021AnConstraints}. Playback buffers have limited efficacy for interactive applications.

End-to-end CC methods function through the interaction of two main mechanisms: The first is \textit{congestion signaling}, which is how congestion signals are generated and sent to the traffic sources. The other mechanism is \textit{rate adaptation}, which is how the traffic sources react to congestion signals. The most crucial measure of success for a CC method is that it avoids congestion collapse, but CC algorithms are also concerned with network performance during regular operation. The performance goal of TCP has traditionally been that all competing flows achieve, on average, a fair share of the total throughput \cite{Jacobson1988CongestionControl} (Though there are different opinions on how to define fair sharing). Recently proposed methods have expanded the performance goals of TCP to include low loss and low queuing delay \cite{Alizadeh2010DataDCTCPb,  Briscoe2021LowArchitecture}.

%We have reviewed two recent survey papers on low latency CC research for mobile networks \cite{Haile2021End-to-endNetworks, Lorincz2021APerspectives}. A striking feature of the surveys is that all of the surveyed work on low-latency CC has focused on algorithms that \textit{converge to} a steady-state where queuing delays are small. Focusing exclusively on steady-state ignores latency performance during the transient period, even though transient latency spikes can affect application performance and user experience. Transient behavior is especially significant in wireless networks because wireless links are typically less stable. More unstable links mean transient behavior makes up a more significant part of regular operation. None of the surveys mention the issue of transient latency spikes.

The original way of signaling congestion in the Internet was to drop packets that arrive at a full queue \cite{Jacobson1988CongestionControl}. Since Van Jacobson's seminal paper, many improvements to congestion signaling have been proposed, tested, and deployed in the Internet. Nichols and Van Jacobson proposed \textit{CoDel}, an algorithm that improves congestion signaling by dropping head-of-line packets when queuing delay exceeds a threshold for too long \cite{Nichols2012ControllingDelay}. In 2010, Alizadeh et. al. introduced Data Center TCP (DCTCP) \cite{Alizadeh2010DataDCTCPb}. DCTCP uses the Explicit Congestion Notification (ECN) bit in the IP header to signal congestion. Because the performance cost of ECN congestion signaling is small compared to dropping packets, ECN makes it feasible to signal congestion earlier and more often. Earlier and more frequent congestion signals allow for changes to the rate adaptation mechanism, and DCTCP takes advantage by making smaller and more frequent adjustments to the transmission rate. DCTCP results in much better latency and packet loss performance than preceding TCP methods, but extensive use of DCTCP is so far limited to data centers. Several adaptations of DCTCP have been suggested with the intention of scaling DCTCP to the entire Internet. These include Low Latency Low Loss Scalable Throughput (L4S) and Some Congestion Experienced (SCE), proposed by Briscoe et al. \cite{Briscoe2021LowArchitecture} and Morton et al. \cite{Morton2021TheCodepoint} respectively.
Bottleneck bandwidth and round-trip propagation time (BBR) is another approach to Internet-scale low latency CC, proposed in 2016 by Cardwell et al. \cite{Cardwell2016BBR:Control}. BBR has similar goals to L4S/SCE, but the BBR algorithm does not use the ECN bit. BBR uses a model of the end-to-end link capacity along with continuous monitoring of the round-trip time (RTT) to detect congestion events. BBR attempts to learn the capacity of the bottleneck interface in the end-to-end path and then adapts its rate to avoid overloading the bottleneck. The innovations of BBR are, therefore, both in the rate adaptation mechanism (estimates of bottleneck bandwidth and RTT instead of reactions to losses or ECN-markings) and congestion signaling (using delays above the minimum instead of losses/markings). DCTCP, L4S, SCE, and BBR are all capacity-seeking because they aim to use close to 100\% of the available bandwidth of the bottleneck link.

Achieving 100\% of the bottleneck capacity is more challenging when the bottleneck capacity varies over time. As examples of the state of the art of low-latency CC over variable-capacity links, we cite two recent surveys of end-to-end CC algorithms for 4G/5G networks by Haile et al. \cite{Haile2021End-to-endNetworks}, and Lorincz et al. \cite{Lorincz2021APerspectives}. Both surveys conclude that nobody has been able to solve the problem of reliable low-latency CC for cellular networks without sacrificing utilization. The authors cite significant capacity variations as one of the main reasons this problem is so hard. Both surveys argue that the magnitude of capacity variations will likely increase with the deployment of 5G, especially over mmWave links. Srivastava et al. \cite{Srivastava2020AnLinks} measure the performance of TCP BBR \cite{Cardwell2016BBR:Control} and TCP Prague (L4S) \cite{Briscoe2021LowArchitecture} over mmWave links where capacity varies over time. Their results show large transient latency spikes when capacity drops for all the congestion control algorithms tested. The reason nobody has been able to demonstrate low-latency capacity-seeking CC over 4G or 5G networks may be that the problem cannot be solved without reducing the variability of the links. Our results provide some evidence for this claim.

%Recent years have shown much progress for low-latency congestion control algorithms. Can we expect continued improvements in the coming years, or are we approaching theoretical limits? A search for works on the theoretical limitations of end-to-end congestion control revealed the literature to be surprisingly sparse. Nevertheless, there are some notable works. Jaffe \cite{Jaffe1981FlowNondecentralizable} proved that it is impossible for a decentralized algorithm to optimize several variations on the \textit{network power} metric, defined as variations on the ratio between aggregates of throughput and latency. Jaffe \cite{Jaffe1981BottleneckControl} also showed that a decentralized algorithm can optimize total throughput while maintaining fairness (fairness defined as equal sharing of bottleneck bandwidth), given that each traffic source knows the capacity of the bottleneck link on its end-to-end path. The works of Jaffe focus on whether or not CC algorithms converge under the assumption that network links have a constant rate. In contrast, our result quantifies the theoretically optimal (minimal) magnitude of transient latency spikes when link capacity is dynamic.

We propose two novel contributions in this paper. First, we describe a theoretically optimal capacity-seeking end-to-end CC algorithm. To be clear, we are not suggesting a new algorithm for deployment in the Internet. The algorithm we describe is intended as a best-case CC algorithm with theoretically optimal congestion signaling. Analyzing an optimal algorithm ensures that our conclusions apply to all capacity-seeking end-to-end CC methods. The second contribution is to show that the algorithm we analyze creates considerable transient delays under very general conditions. Because we analyze a theoretically optimal algorithm, we can conclude that it is impossible for any practical algorithm to perform better under the same conditions. Our result, therefore, maps part of the border between possible and impossible performance goals for capacity-seeking end-to-end congestion control algorithms.

\section{Analysis}
\label{sec:analysis}
\noindent This section describes and analyzes a theoretically optimal capacity-seeking end-to-end CC algorithm to establish a bound on the performance of this class of algorithms in general. To be clear, we are not proposing a novel CC algorithm. Instead, we describe and analyze a theoretically optimal algorithm (which is not implementable in practice) and draw conclusions concerning all capacity-seeking end-to-end congestion control algorithms.

% \begin{figure}[t]
%     \centering
%     \includegraphics[width=\linewidth]{figures/latency_end_to_end_cc.pdf}
%     \caption{An end-to-end path in the Internet}
%     \label{fig:path}
% \end{figure}
\begin{figure}
     \centering
     \subfloat[ACK-based congestion control saturating an interface $X$]{\label{fig:saturatedCCA}
        \includegraphics[width=0.90\linewidth]{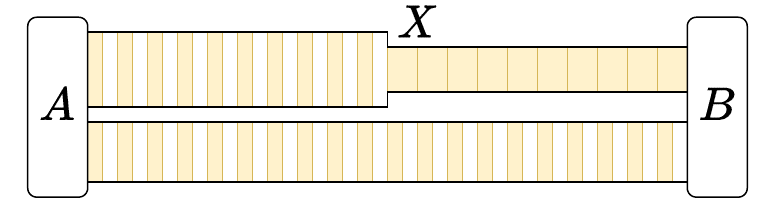}} \\
     %\hfill
     \subfloat[Packet flow after capacity is reduced at interface $X$ for a flow controlled by ACK-based end-to-end congestion control]{\label{fig:stepchangeCCA}
        \includegraphics[width=0.90\linewidth]{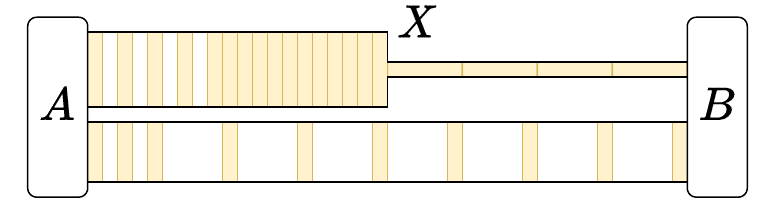}} \\
     %\hfill
     \subfloat[Packet flow after capacity is reduced at interface $X$ for a flow controlled with our theoretically optimal congestion controller]{\label{fig:optimalCCA}
        \includegraphics[width=0.90\linewidth]{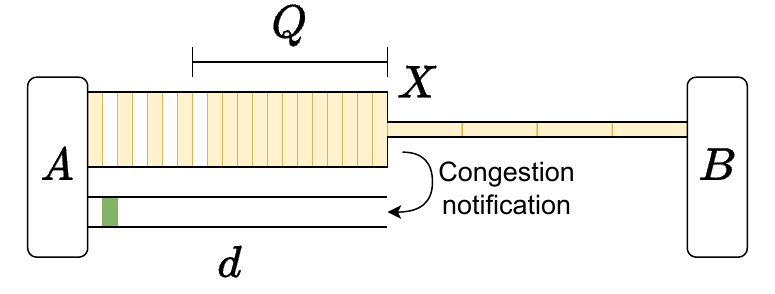}}
    \caption{A visualization of end-to-end congestion control}
    \label{fig:three graphs}
\end{figure}

Consider a sender, $A$, transmitting a capacity-seeking flow to a receiver $B$ through a bottleneck interface $X$. $A$ aims to use all of the available capacity of interface $X$; therefore, we assume that interface $X$ is fully loaded with traffic from $A$.
Our analysis centers on what happens if the capacity of $X$ is suddenly and unpredictably reduced.
We make several simplifying assumptions to ensure our conclusions are as general as possible. By over-estimating the capabilities of the end-to-end controller, we ensure our results can safely be regarded as best-case performance for any real-world implementation. In other words, we are modeling the most favorable case for end-to-end CC, thereby producing a bound on the performance of capacity-seeking end-to-end CC algorithms.

In a typical real-world network, the interface $X$ does not know how to talk directly to $A$. Congestion signals therefore typically travel via $B$ before reaching $A$ (see figures \ref{fig:three graphs}(a) and \ref{fig:three graphs}(b)). Both TCP and QUIC do congestion signaling via the receiver. In our analysis, we assume $X$ has a direct link to $A$ with constant delay $d_{signal}$, as shown in figure \ref{fig:three graphs}(c). A direct and constant latency link represents the best-case scenario for any congestion signaling method. The latency from $A$ to $X$ is denoted $d_{flight}$. In the general case, $d_{flight}$ and $d_{signal}$ need not be equal. In case the congestion signal travels via $B$, then $d = d_{flight} + d_{signal}$ is equal to the round-trip time.
We also assume the queue at $X$ is empty at time $t_{0}$. It is possible for the queue at $X$ to be empty and for the interface $X$ to be fully loaded if we assume packet arrival and departures are perfectly in sync and happen simultaneously. This is the ideal case of zero queuing and 100\% throughput that several low-latency congestion control methods, such as BBR \cite{Cardwell2016BBR:Control} and DCTCP \cite{Alizadeh2010DataDCTCPb}, attempt to approximate.

To summarize, we make the following simplifying assumptions about the end-to-end CC method:

\begin{itemize}
    \item A congestion signal including information about the new capacity is created at $X$ the instant capacity is reduced.
    \item The congestion signal travels from $X$ to $A$ in the shortest possible time $d_{signal}$
    \item $d_{flight}$ is the time it takes a packet from $A$ to reach $X$
    \item On receiving the signal, $A$ reacts immediately by reducing the load to exactly match the new capacity at $X$
    \item The queue at $X$ is empty when the capacity drops
    \item There is infinite buffer space at $X$ so no traffic is lost
    \item For simplicity, we ignore the fact that $A$ must reduce load below the capacity at $X$ to allow the queue to empty
\end{itemize}

\noindent \textbf{Step change in capacity}

\noindent Assume $X$ has an original capacity of $C_{0}$ Mbit/s. The capacity of $X$ is now instantaneously reduced by a factor $1/r$, such that the new capacity $C_{1}$ is given by equation (\ref{eq:capreduction}).

\begin{equation}
    \label{eq:capreduction}
    C_{1} \text{ Mbit/s} = \frac{C_{0} \text{ Mbit/s}}{r}
\end{equation}

% \begin{table}[t]
%     \centering
%     \begin{tabular}{l|c}
%         & $d$ \\
%         \hline
%         Speed of light Dublin - New York & 17 ms\\
%         Theoretically Optimal LEO Satellite Dublin - New York \cite{Chaudhry2021OpticalPerspective} & 20.07 ms\\
%         Theoretical Optical Terrestrial Cable Dublin - New York \cite{Chaudhry2021OpticalPerspective} & 25.07 ms\\
%         Speed of light around the equator & 133 ms
%     \end{tabular}
%     \caption{Dublin to New York one-way delays}
%     \label{tab:dublinnewyork}
% \end{table}

When the capacity of $X$ is reduced at time $t_{0}$, $X$ immediately sends a congestion signal towards $A$. Figure \ref{fig:optimalCCA} shows the situation moments before $A$ receives the congestion signal (shown in green). The earliest time $A$ can know about the capacity reduction is after $d_{signal}$ milliseconds, where $d_{signal}$ is the minimum delay from $X$ to $A$. The reduction in capacity implies that the per-packet processing time at $X$ increases by a factor $r$. It now takes $r$ milliseconds for $X$ to transmit the amount of data that used to take one millisecond before the capacity reduction. Therefore, the arrival rate at $X$ is now $r$ milliseconds of traffic every millisecond. A queue is now building up at $X$ at a rate of $(r-1)$ milliseconds of queuing latency every millisecond. Because $d_{signal}$ milliseconds pass before $A$ receives the congestion signal and reduces the rate, and because $d_{flight}$ of data was already in flight when capacity dropped, the size of the queue at $X$ will grow by $(r - 1) * (d_{signal} + d_{flight})$ ms. Since we assume the queue is empty at time $t_{0}$, the queue delay peaks at $(r-1)*d$ ms where $d = d_{signal} + d_{flight}$.

\begin{figure}
     \centering
    \includegraphics[width=0.97\linewidth]{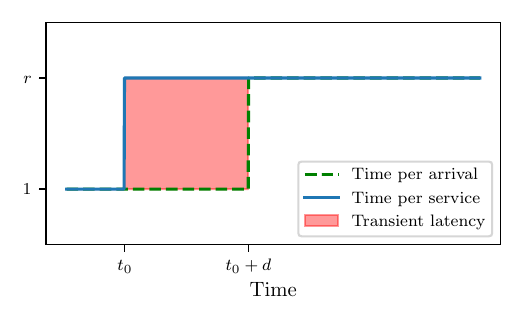}
    \caption{Per-packet time relative to the processing time at the original capacity $C_{0}$}
    \label{fig:instantdrop}
\end{figure}

The peak transient delay, $Q$, produced as a result of an instantaneous capacity reduction is given by equation (\ref{eq:mindelay}). Figure \ref{fig:instantdrop} illustrates how the queue delay can be visualized as the area spanned out by $d$ and $(r-1)$. Notice that $Q$ is independent of the throughput of $X$.

\begin{equation}
    \label{eq:mindelay}
    Q = (r - 1) * (d_{signal} + d_{flight}) = (r - 1) * d
\end{equation}

\begin{figure}
    \centering
    \includegraphics{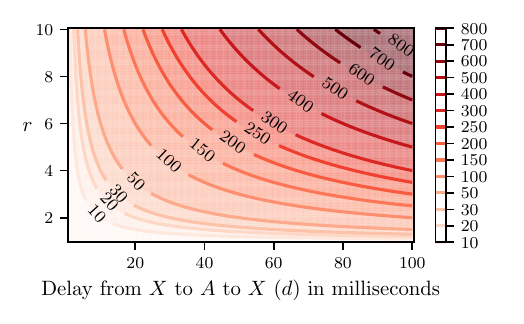}
    \caption{Minimum transient queuing delay (in milliseconds) with end-to-end congestion control and no packet loss}
    \label{fig:transientdelay}
\end{figure}

Figure \ref{fig:transientdelay} shows how $Q$ scales as $d$ and $r$ changes. The value of $Q$ grows to \textit{several hundred milliseconds} for the selected ranges of $d$ and $r$. If the queue at $X$ has limited capacity, contrary to our assumptions, then any overflow will be translated into packet loss.
\vspace{1mm}

\noindent \textbf{Packet loss}

\noindent If buffer space is limited, traffic can be lost instead of contributing to the size of the latency spike. There are different ways to limit the queue size, and here we explore the specific case where the queue is limited by a maximum queuing delay. If the queue is limited by the time traffic has been waiting in the queue, the queue delay has an upper bound $\Bar{q}$. The part of the latency spike above $\Bar{q}$ is lost. We can calculate the percentage of traffic lost in the interval $[t_{0}, t_{0}+d]$, $l_{\Bar{q}}$, using equation (\ref{eq:timelimitloss}).

\begin{equation}
\label{eq:timelimitloss}
    l_{\Bar{q}} = \frac{Q - \Bar{q}}{Q}
\end{equation}

The value of $\Bar{q}$ determine how $X$ handles the trade-off between the size of the latency spike and the amount of packet loss for this case of limited buffer space. Several well-tested algorithms, such as CoDel \cite{Nichols2012ControllingDelay} and PIE \cite{Pan2013PIE:Problem} fall somewhere in-between setting a maximum queue delay $\Bar{q}$ and just letting the queue grow to the value $Q$ given by equation (\ref{eq:mindelay}). Equations (\ref{eq:mindelay}) and (\ref{eq:timelimitloss}) can not directly predict the performance of these algorithms. However, the equations enable us to compute the bounding cases of how much delay we get if we insist on no loss (Eq. (\ref{eq:mindelay})) and how much loss we get if the delay is strictly bounded (Eq. (\ref{eq:timelimitloss})). The actual performance of CoDel, PIE, and other active queue management algorithms aiming to limit queuing delay will fall somewhere between these two by trading some reduction in peak latency for some increase in packet loss.

\section{Experimental evaluation}
\begin{figure}
    \centering
    \begin{tikzpicture}[node distance=0cm]
    \tikzstyle{process} = [rectangle, minimum width=2.2cm, minimum height=2.4cm, text centered, draw=black, fill=orange!30]
    \tikzstyle{arrow} = [thick,->,>=stealth]
    \node (client) [process, align=center] {Client};
    \node (server) [process, right of=client, xshift=6cm, align=center, fill=green!30] {Server};

    \path (client.east) -- (client.south east) coordinate[pos=0.5] (a1);
    \path (server.west) -- (server.south west) coordinate[pos=0.5] (b1);
    \path (client.east) -- (client.north east) coordinate[pos=0.5] (a2);
    \path (server.west) -- (server.north west) coordinate[pos=0.5] (b2);
    \node (HTB) [left of=b1, xshift=-0.6cm] {HTB};
    \node (CoDel) [left of=HTB, xshift=-1.2cm] {CoDel};
    \node (netem) [left of=b2, xshift=-0.6cm] {netem};
    \draw (HTB) circle [radius=0.6cm];
    \draw (CoDel) circle [radius=0.6cm];
    \draw (netem) circle [radius=0.6cm];

    \draw [arrow] (a1) -- node[anchor=north] {Data} (CoDel);
    \draw [arrow] (netem)+(-0.6, 0) -- node[anchor=south] {ACKs} (a2);
    \draw [>=latex,->,>=stealth,font=\scriptsize, thick] (b1) -- +(0.4,0) |- (b2);
    
    \end{tikzpicture}
    \caption{The testbed configuration. HTB limits the rate, Netem delays the congestion signal, and CoDel signals congestion using ECN markings.}
    \label{fig:testbed}
\end{figure}
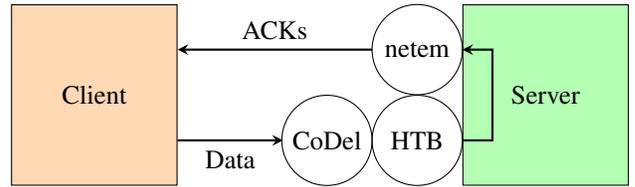

Does the lower bound on latency spike size hold in the real world? To investigate, we built a simple testbed consisting of two Raspberry Pi 4B machines and a network switch\footnote{A repository containing our tests, raw data, and scripts is available here: https://github.com/bjornite/thesis-tcp-testbed}. Using the Linux tool \textit{netem} and the hierarchical token bucket (HTB) and CoDel queuing disciplines (qdiscs), we emulate a link with configurable rate and delay. Rate is limited on the server ingress, and delay is added on the server egress. The testbed setup is depicted in figure \ref{fig:testbed}. Idle round-trip delay was measured to be consistently below 1 ms, so we assume transmission delays can be ignored in this experiment. We disable TCP offloading on all active network interfaces on both the client and the server. Latency is recorded using TCP round-trip time statistics through Flent.

\begin{figure}
    \centering
    \includegraphics{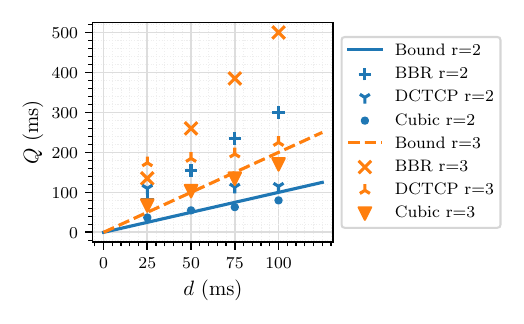}
    \caption{Peak transient queuing delay. The lines show the analytical lower bound of equation \ref{eq:mindelay}, and markers show experimental results for the Cubic, BBR and DCTCP congestion control algorithms.}
    \label{fig:experiment}
\end{figure}

We ran the tcp\_1up test in Flent \cite{Hiland-Jrgensen2017Flent:Tester} to generate a single TCP upload from client to server. The link rate was set to an initial value of 60 Mbit/s. The test lasts for 60 seconds, and at the 30 second mark of each experiment, the rate was reduced by a factor $1/r$. We ran the experiment for BBR, Cubic and DCTCP using several different values of $r$ and $d$. We record the difference between peak delay during the transient period, defined as 29s-34s experiment runtime, and the value of $d$. Figure \ref{fig:experiment} shows the results. 

The only cases where measured values are below the bound are when Cubic is combined with a value of $d > 50ms$. In these cases, Cubic does not saturate the link. This reduces the effective value of $r$, which explains why Cubic appears to do better than the bound in these instances. When we correct for the lack of link saturation, the bound does hold for these cases as well. The measured values are above the bound for all cases using BBR and DCTCP. We can therefore conclude that the bound does in fact seem to hold for Cubic, the most widely deployed TCP congestion control, as well as for newer algorithms such as BBR and DCTCP.

\section{Discussion}
\label{sec:discussion}

\noindent We have derived an expression for the minimum size of the latency spike when capacity suddenly drops. This lower bound is a function of the magnitude of the capacity drop, $r$, and delay, $d$. Our experiments show that the lower bound holds for BBR \cite{Cardwell2016BBR:Control} in a simple testbed. This section illustrates how our analysis can inform decisions in real-world networks.
\vspace{1mm}

\noindent \textbf{The data center}

\noindent Capacity-seeking end-to-end CC can achieve low latency and low loss in data centers, as demonstrated by DCTCP \cite{Alizadeh2010DataDCTCPb}. Because of the short physical distances within a data center, signaling and in-flight delay $d$ can be very small. As long as $r$ is kept small enough, transient queuing delays can be maintained at a manageable level. One might assume that DCTCP can somehow be made to work equally well for the Internet. However, as shown, this is not the case for links with variable capacity.
\vspace{1mm}

\noindent \textbf{Fair queuing and round-robin schedulers}

\noindent On good quality fiber or copper links, variations in capacity are most likely due to competing traffic. We consider here the case of a single fair-queuing interface shared between multiple users. With fair queuing, each user (or other traffic classification) has its own queue, and the per-user queues are serviced in a round-robin fashion. In this scenario, $r$ depends on how many other users are connected and how many new users connect simultaneously. If a user is alone on the shared link, and then one more user connects, the first user will see a drop in capacity of $r = 2$. If two users join simultaneously, then $r = 3$.
\vspace{1mm}

\noindent \textbf{Magnitude of capacity changes in wireless technologies}

\noindent In wireless technologies, the link rate is often dynamically adapted based on channel conditions.
Channel conditions depend on many factors, such as distance, occlusions, and reflections. Because there are many sources of variation in channel capacity, the link rate can change frequently, rapidly, and in large jumps. In addition to varying channel conditions, each wireless channel may be used by more than one radio. The resulting contention for the wireless channel is another source of sudden capacity variations.

Due to the many sources of capacity variations in typical wireless networks, we can not rely on $r$ being always less than 10 for a typical WiFi or 5G connection. The signaling delay $d$ solely due to light-speed delay in the Internet is often measured in tens of milliseconds. Therefore, based on the results shown in figure \ref{fig:transientdelay}, we can conclude that capacity-seeking end-to-end CC can not reliably deliver low latency over typical WiFi and 5G wireless links. We believe this problem will likely persist with new WiFi standards and 6G mobile networks.
\vspace{1mm}

\noindent \textbf{Multiple flows with different signaling delay}

\noindent What happens when multiple flows with different values of $d$ share the same FIFO bottleneck interface? If all of the assumptions listed in section \ref{sec:analysis} are true, the peak queuing delay will be the weighted sum of $Q$ (see Eq. (\ref{eq:mindelay})) for each of the flows. The weights will be the relative bandwidth share of each flow at $t_{0}$.
\vspace{1mm}

\noindent \textbf{Sketching solutions}

\noindent It is beyond the scope of this paper to propose a solution to the lag spike issue. Nevertheless, our analysis may serve as a framework for reasoning about the problem. A solution must bound the values of $d$ or $r$ for capacity-seeking end-to-end flows that require consistent low latency. To illustrate the usefulness of this model as a reasoning tool, we list a few methods that might be part of a solution:

\begin{itemize}
    \item Under-utilization of the link reduces $r$.
    \item When $r$ is partially due to competing traffic, ramping up competing traffic sources more gradually will reduce $Q$.
    \item If capacity drops can be predicted, the effect of reducing $d_{signal}$ can be achieved by signaling congestion early. Zhuge is an example of this \cite{Meng2022AchievingLoop}.
    \item Content delivery networks (CDNs) place servers close to clients and reduces $d_{signal}$ and $d_{flight}$.
    \item Prioritizing congestion signals (often this means ACKs) over other traffic in the reverse path can improve $d_{signal}$.
    \item Dropping head-of-line packets instead of last-in-line packets reduces $d_{signal}$. CoDel is an example of this \cite{Nichols2012ControllingDelay}.
    \item Reducing the interaction effects of different users or flows on each other can reduce $r$ (and $d_{signal}$ if congestion signals are delayed by other traffic). Fair queuing is an example of this \cite{Hiland-Jrgensen2018PieceGateways}.
    \item Forwarding information at many levels of fidelity can move the capacity adaptation decision closer to the bottleneck interface. In theory, this can reduce $d$ to almost zero but comes at the cost of more resource consumption upstream of the bottleneck.
\end{itemize}

%\balance
\section{Conclusion}
\label{sec:conclusion}

\noindent In a worldwide network where large and rapid drops in capacity are likely to happen, capacity-seeking end-to-end congestion control cannot avoid spikes in latency or packet loss. We have shown how a lower bound on the latency spikes is determined by the signaling delay of congestion notifications and the scale of the capacity variations. The simplicity of our assumptions makes the analysis very generally applicable. The results are valid for all capacity-seeking end-to-end congestion control methods, including all versions of TCP, QUIC, and adaptive bit-rate algorithms for voice and video. Research on improving latency and packet loss in the Internet must account for this limitation of capacity-seeking end-to-end congestion control. Our results highlight the need for mitigation methods such as fair queuing. Because of the large and frequent capacity variations in wireless communication, our analysis is especially relevant for wireless technologies, including WiFi and 5G. Future work will evaluate more TCP versions in the testbed, measure how CoDel and PIE trade latency for packet loss, and analyze the trade-offs of end-to-end congestion control when link capacity varies in more complicated ways.

%\balance

\bibliographystyle{IEEEtran}
\bibliography{IEEEabrv, references}  %%% Remove comment to use the external .bib file (using bibtex).
%% and comment out the ``thebibliography'' section.

\end{document}